\documentclass[boxit]{JAC2003}

\usepackage{graphicx}
\usepackage{booktabs}
\usepackage{verbatim}
\usepackage{subcaption}
\begin{document}

\title{Development of a Hybrid Power Supply and RF Transmission Line for SANAEM RFQ Accelerator}
\author{S. Ogur \thanks{salim.ogur@boun.edu.tr}, Bogazici University, Istanbul, Turkey \\ G. Turemen, Ankara University, Ankara \\ F. Ahiska, EPROM Electronic Project \& Microwave Ind. and Trade Ltd. Co., Ankara, Turkey \\  A. Alacakir, TAEK-SANAEM, Ankara, Turkey  \\
G. Unel, UCI, Irvine, California, USA\\}

\maketitle

\begin{abstract}
SANAEM Project Prometheus (SPP) has been building a proton beamline at MeV range. Its  proton source, two solenoids, and a low energy diagnostic box have been already manufactured and installed. These are going to be followed by a 4-vane RFQ to be powered by two stage PSU. The first stage is a custom-built solid state amplifier providing 6 kW at 352.2 MHz operating frequency. The second stage, employing TH 595 tetrodes from Thales, will amplify this input to 160 kW in a short pulsed mode.  The power transfer to the RFQ will be achieved by the means of a number of  WR2300 full and half height waveguides, 3 1/8" rigid coaxial cables, joined by appropriate adapters and converters and by a custom design circulator. This paper summarizes the experience acquired during
the design and the production of these components.
\end{abstract}

\section{Introduction}
\begin{figure}
\begin{centering}
\includegraphics[width=0.95\columnwidth]{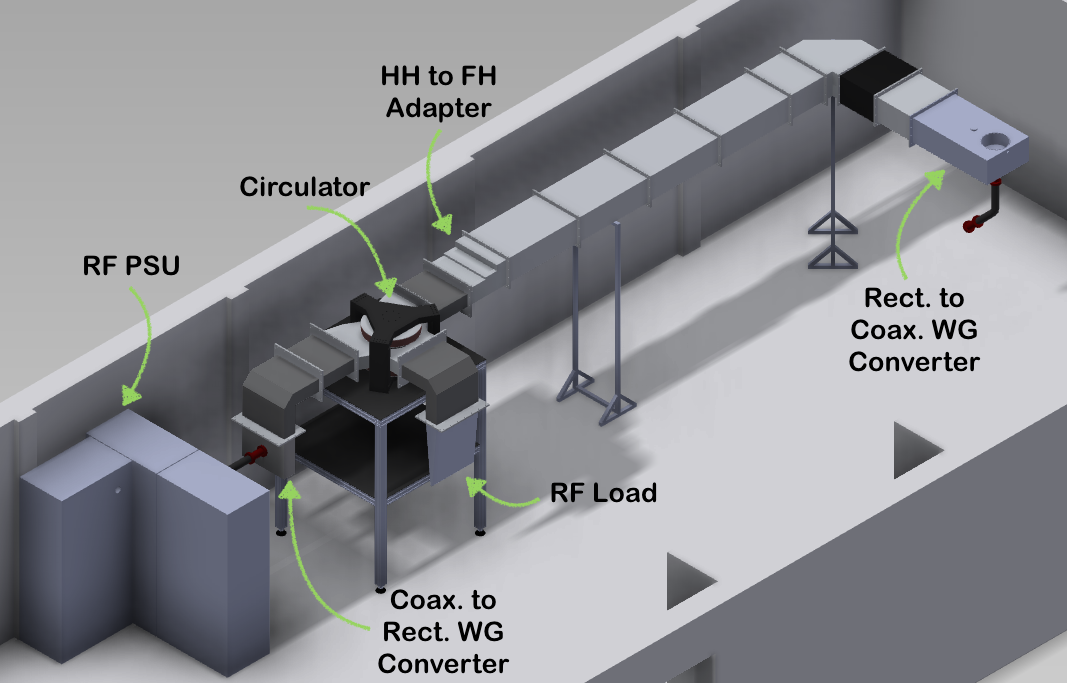}
\par\end{centering}
\protect\caption{The overview of the SPP RF transmission line. \label{fig:overview}}
\end{figure}

SANAEM, Saraykoy Nuclear Research and Training Center of Turkish Atomic Energy Authority in Ankara, has begun to build a Proton Linac, under the name of SPP, with a modest target: aiming to get at least 1 mA of current at the energy of 1.3 MeV \cite{overall}. This low-energy proton linac is also going to serve as a domestic know-how build-up instrument by which young scientists and engineers, as well as the local industry, will get acquainted with accelerator technology. The RF power section of this linac consists of three main parts: the Power Supply Unit (PSU), a circulator, and the power transmission line including appropriate waveguide adapters and converters.
The RF transmission system is to feed the 4-vane RFQ with the design parameters presented in elsewhere \cite{overall}, \cite{RFQ}. The  power requirement for the SPP RFQ is calculated by taking into account
the ohmic losses on the RFQ walls (60 kW obtained after computer simulations) together with the transmission and coupling losses (30 kW a worst case scenario estimation). Folding in a safety factor of 1.3, one finds 120 kW as the required RF power under the assumption of continuous operation. SSP will initially operate with a  duty factor less than 3\% to reduce the requirements on the RF PSU. The entire RF power supply and transmission system is shown in Fig.~\ref{fig:overview}.

\section{POWER TRANSMISSION LINE}
\subsection*{Power Supply Unit}
\begin{figure} 
\begin{centering}
\includegraphics[width=0.95\columnwidth]{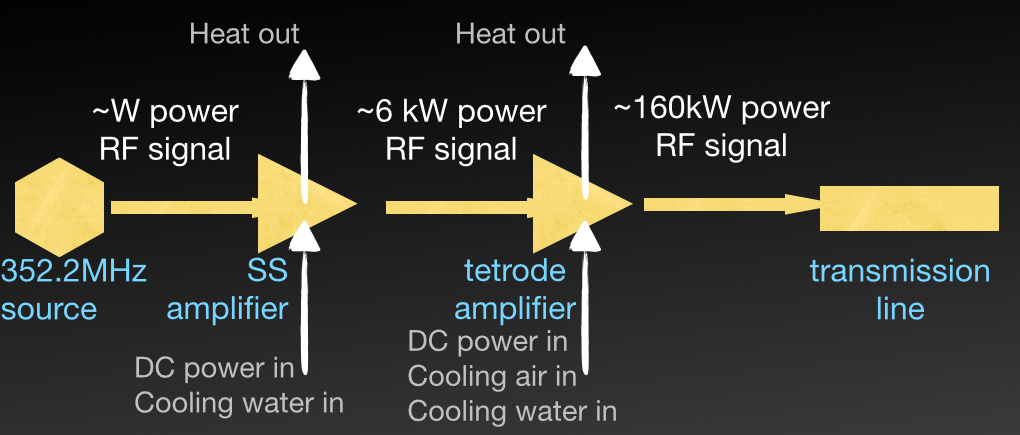}
\par\end{centering}
\protect\caption{SPP PSU design.\label{fig:SPP-PSU-design}}
\end{figure}

 The operating frequency of the SPP RFQ is 352.2 MHz. This operating frequency is chosen to be compatible with similar machines in Europe and to take advantage of already available RF amplifier market. The accelerator is intended to operate at a maximum duty factor of about 3\%. To fulfill these requirements, two stage amplification has been selected, as summarized in Fig.~\ref{fig:SPP-PSU-design}. The power supply  is designed by the SPP-team, and manufactured by a local company \cite{eprom}.

\begin{figure} 
\begin{centering}
\includegraphics[width=0.95\columnwidth]{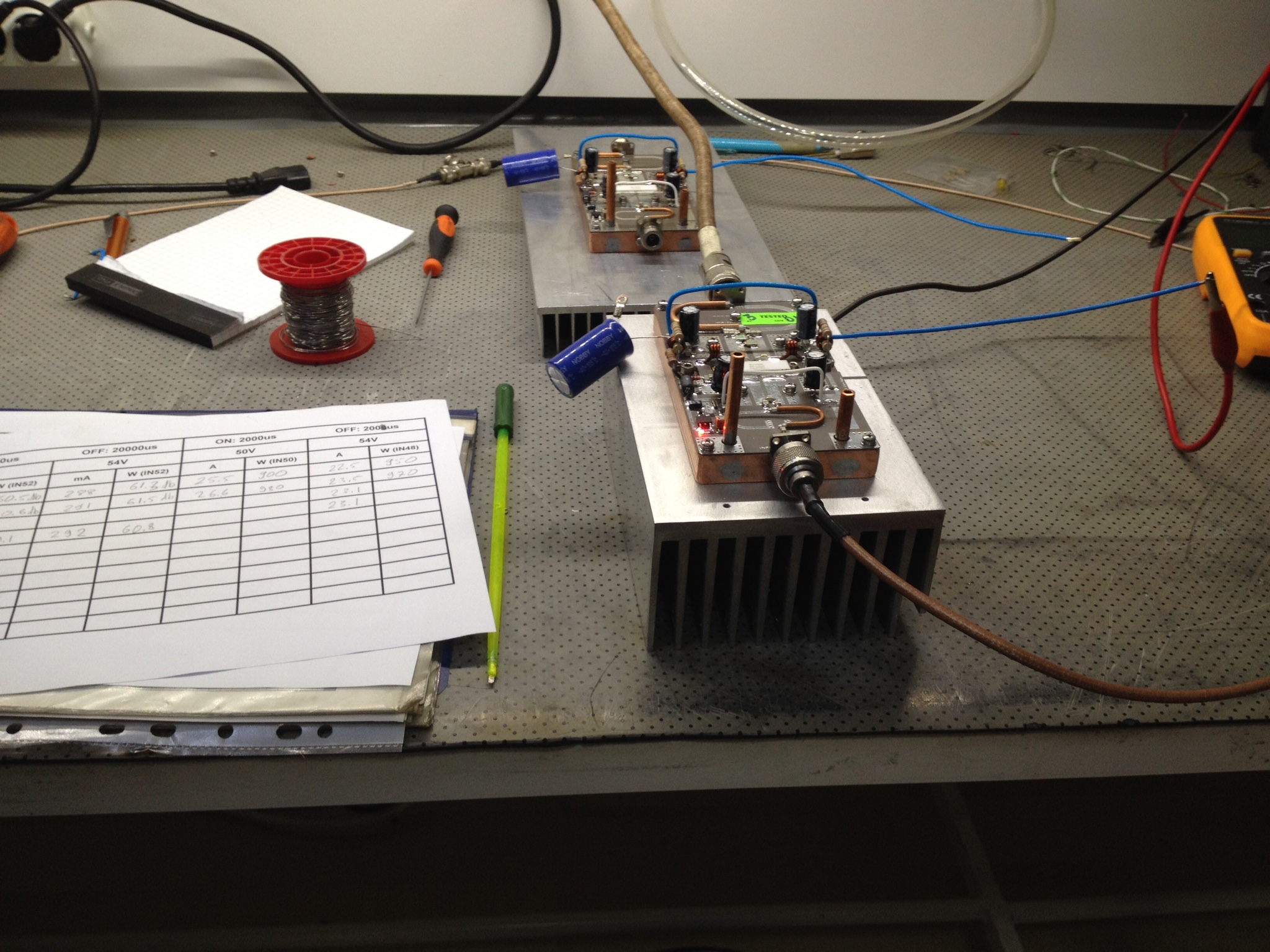}
\par\end{centering}
\protect\caption{SPP SSA test bench.\label{fig:SS-test}}
\end{figure}

\begin{figure}
\begin{centering}
\includegraphics[width=0.9\columnwidth]{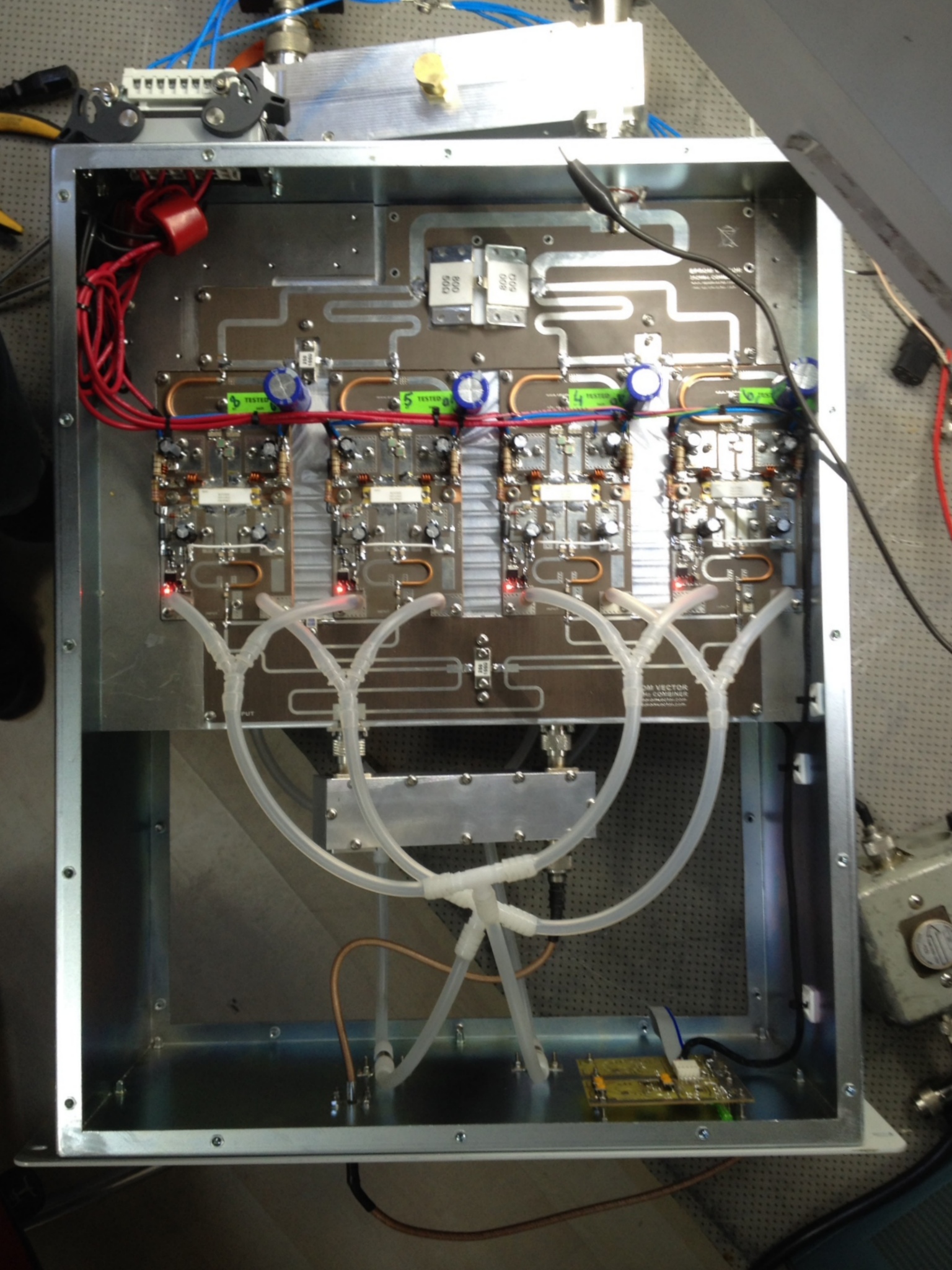}
\par\end{centering}
\protect\caption{SPP PSU solid state amplifier combiner.\label{fig:SPP-SSA-8combi}}
\end{figure}

The construction of the first stage of the PSU started around the BLF-578XR high power transistor \cite{BLF}. The water cooled solid state amplifier circuit designed around this transistor provides about 1.7 kW of RF power for a pulse length shorter than 500 milliseconds, with an amplification of about 18.3 dB. A photo of the SS amplifier unit test bench can be seen in Fig.~\ref{fig:SS-test}.
 The combiner system, consisting of 8 such boards, achieves a total power of 6 kW in continuous mode and over 10 kW in pulsed mode. For this particular application, the output power was limited to about 7 kW. The photo of the amplifier combiner system is shown in Fig.~\ref{fig:SPP-SSA-8combi}.

The second stage of  amplification continues with the TH595 tetrode from Thales\cite{thales} which has a gain around 15 dB. It can operate up to 450 MHz just like its predecessor TH391. The TH595 is to be cooled with both by air convection and water conduction. The improved cooling at the amplifier tube yields higher power dissipation at the anode, resulting in its preference over the TH391.  The TH595 can provide 60 kW power at CW operations and has a maximum output of 200kW at a pulse length of 3.3ms.  Optimizing the input and output RF powers and the duty factor values, the maximum output power was selected as 160kW, corresponding to an amplification of 14.3 dB. Although 160 kW RF power can be provided up to 5ms by the selected tetrode, the current implementation has a user selectable active time ranging from 200 $\mu$s to 2 ms. Therefore the user will be able to further optimize the duty factor and the repetition rate. TH595, together with its cavity is expected to cover all SPP needs while its cost remains considerably less than tetrodes' competitors: the klystrons.  For the final product, the high current power supplies and the control software are already built, using local resources. The whole power supply is being assembled into standard racks  \cite{eprom}.

\subsection{Waveguides, Bends, and Converters }
\begin{figure*}
\begin{centering}
\includegraphics[width=1.95\columnwidth]{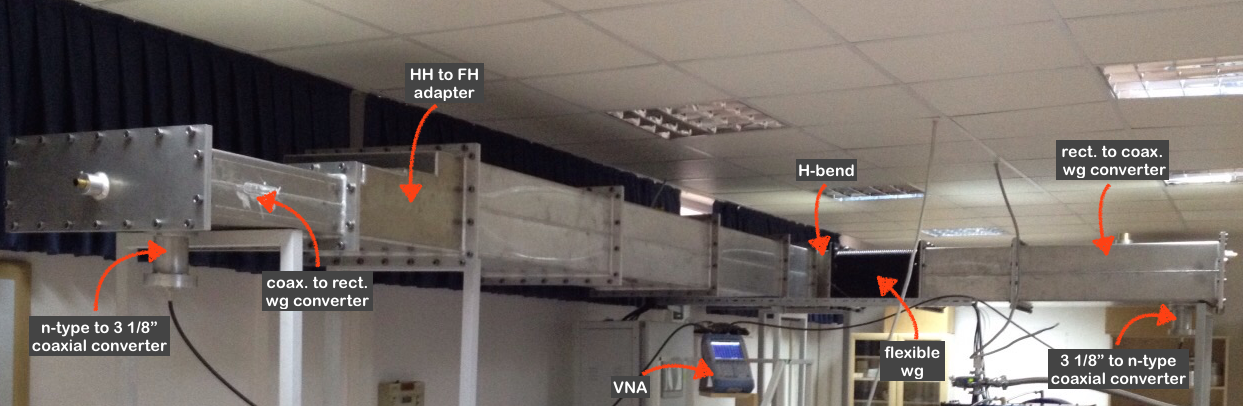}
\par\end{centering}
\protect\caption{Assembled transmission line of the SPP. \label{fig:assembled}}
\end{figure*}

Since SPP operation frequency is 352.2 MHz, WR2300 waveguide (WG) standard, which can convey power up to 700 MWs, is chosen. The RF PSU's coaxial output, is converted to  half-height (HH) WG via a custom designed and manufactured  Waveguide to Coaxial Transition Unit. 
 HH waveguides are going to be utilized around the ports of the HH circulator. One HH E-Bend WG is used to bridge the PSU and the circulator.   Another HH E-bend is used for the circulator port connecting to the RF dump. The  circulator output port leading to the RFQ is converted to full-height (FH) by a HH to FH adapter WG followed by a 3 meters FH transmission line. The RF field  is then rotated 90$^{\circ}$ horizontally by a  FH B-bend. Finally a FH  WG to coaxial converter has been designed and manufactured for delivering power  to the RFQ via a rigid coaxial cable \cite{rigid}.   
All waveguides and converters are designed on CST MWS \cite{cst} to work within an insertion loss below $-$85 dB under PEC assumption. Both HH and FH waveguides, converters and adapters are locally manufactured \cite{kalitek}. 
Only flexible waveguide in the entire setup for SPP RF transmission is bought from a non-local company \cite{mega}. The assembled RF transmission line is shown in Fig.~\ref{fig:assembled}.

    Low level RF tests of the entire RF line as well as of individual sections have been performed using a VNA\cite{network-analyzer}. It was found that the assembled RF transmission line has an insertion loss (S11) below $-$60 dB. 

\subsection{Circulator}

\begin{figure}[htb!]
\begin{center}
\includegraphics[scale=0.33]{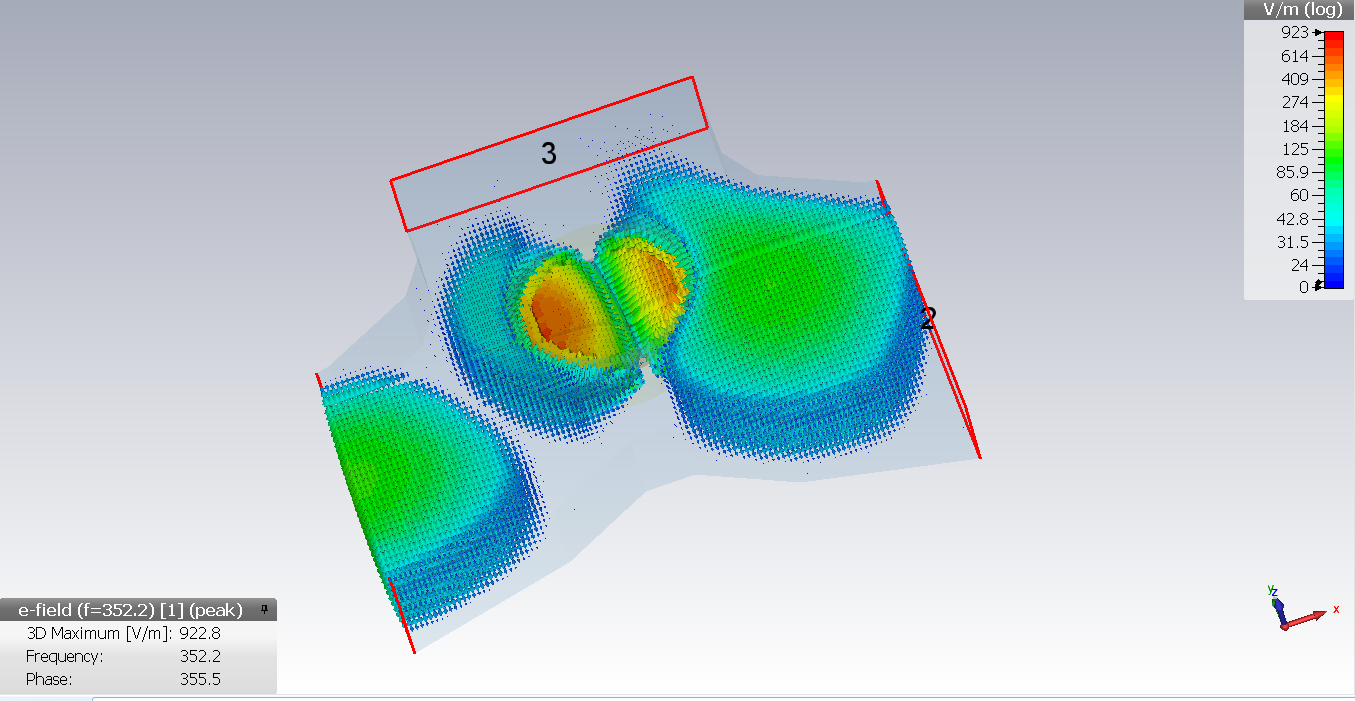}
\caption{Electric field inside the circulator for 352.2 MHz. \label{circ_elec}}
\end{center}
\end{figure}

A non-reciprocal RF device has been designed to aid the power delivery from PSU to RFQ. It transmits RF power from port 1 to port 2 as shown in Fig.~\ref{circ_elec} while the reflected power is dumped out to port 3. Our design is made by using CST MWS, and after lengthy optimization studies,  NG 1600 ferrite disks were ordered from Magnetics Group\cite{magnetics-group}. This material has a saturation magnetization $4\pi M_s$ at 1600 Gauss, its maximum line width is 10 Oersted, its dielectric constant is 14.6 while the loss tangent is below $2\times 10^{-4}$ where all values are given by the provider at 9.4 GHz. The Curie temperature of the ferrite is given as 220 $^\circ$C.  The designed circulator is going to operate in above resonance to have low loss as pointed out by Bosma\cite{Bosma}.

\begin{figure}[htb!]
\centering{}\includegraphics[width=1\columnwidth]{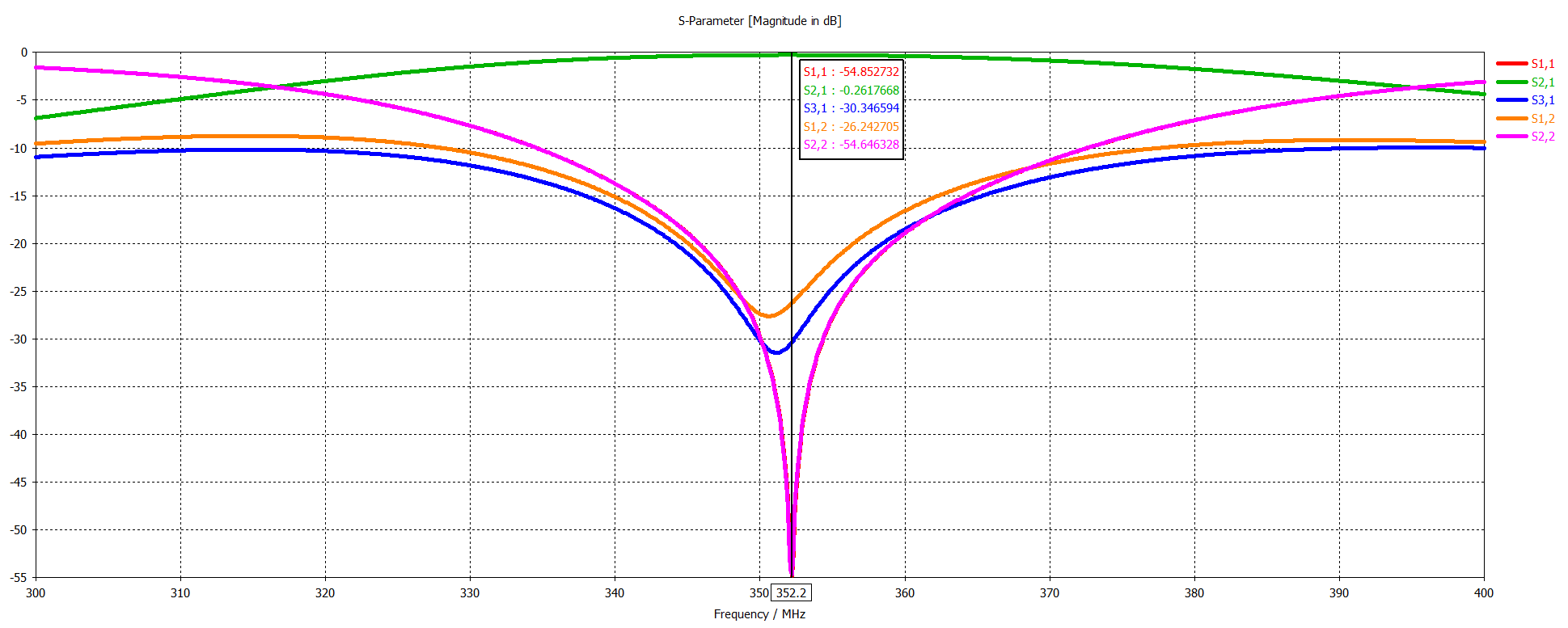} \protect\caption{Performance of the circulator design at 352.2 MHz. \label{fig:circulator_results}}
\end{figure}

\begin{figure}
\centering{}\includegraphics[width=0.9\columnwidth]{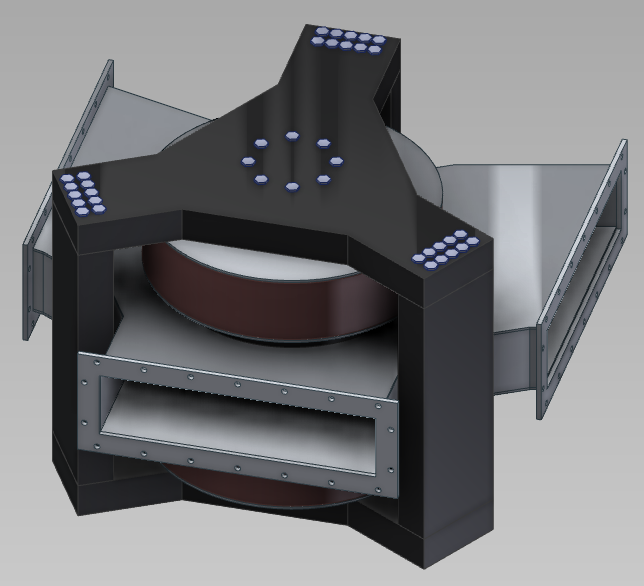} \protect\caption{The 3D CAD model of the designed circulator. \label{circ3d}}
\end{figure}

The mechanical design consisting of two disk-ferrites and attached to aluminum disk on inner sides of the HH WR2300 waveguides. The cooling of ferrites will be both provided by the chilled water  and convectional air cooling through the whole circulator to avoid possible overheat due to transmission losses. The scattering parameters  can be seen in Fig.~\ref{fig:circulator_results}. At the SPP operating frequency of 352.2 MHz, the return loss ($S11$)
is found to be $-$55 dB, the insertion loss ($S21$) is found as $-$0.26 dB and finally the isolation loss
($S31$) is found as $-$30 dB. VSWR is calculated to be about 1.005:1. The engeneering design of the circulator is completed (Fig.\ref{circ3d}) and the production of the components will start soon.

\section{Outlook}
SPP RF PSU will be delivered in Q2 2015 and the transmission line will be subjected to high power RF tests in Q3 2015. The entire RF line needs to be operational by Q4 2015. The first protons are expected to be accelerated by the SPP RFQ by Q1 2016.

\section*{Acknowledgments}

\indent

The authors are grateful to A. Tanrikut for useful comments. This
study is supported by TAEK with a project No.~A4.H4.P1.

\end{document}